\documentstyle[12pt]{article}
\begin{document}
\Large Hilbert C*-systems for actions of the circle group

\vspace{0.5cm}

\large H. Baumgaertel 

\normalsize Erwin Schroedinger International Institute of 
Mathematical Physics, 

Wien, Austria

\vspace{0.1cm}

permanent address:

Institut fuer Mathematik

Universitaet Potsdam

Am Neuen Palais 10

D-14451 Potsdam, Germany

\vspace{0.3cm}

\large A. L. Carey

\normalsize Department of Pure Mathematics

University of Adelaide

Adelaide, Australia

\sloppy

\abstract{
The paper contains constructions of Hilbert systems for the action
of the circle group $T$
using subgroups of implementable Bogoljubov unitaries w.r.t. Fock
representations of the Fermion algebra for suitable data of the selfdual
framework: ${\cal H}$ is the reference 
Hilbert space, $\Gamma$ the conjugation
and $P$ a
basis projection
on ${\cal H}.$ 
The group 
$C(\mbox{spec}\,{\cal Z}\rightarrow T)$
of $T$-valued functions on $\mbox{spec}\,{\cal Z}$ 
turns out to be isomorphic to the stabilizer of ${\cal A}$.
In particular, examples are presented where the center
${\cal Z}$
of the fixed point algebra ${\cal A}$
can be calculated explicitly.\footnote{2000 Mathematics 
Subject Classification 46L60, 81R10}}

\section{Introduction}

It is well-known that the DHR-selection criterion in
 superselection theory
leads to a (global) symmetry group ${\cal G}$ of the theory, acting on the
so-called field algebra.
This group ${\cal G}$ is compact.
These C*-systems 
consisting of
the field algebra, together with the automorphism group
 ${\cal G}$, are called 
{\em Hilbert systems}
(for a precise definition see below)
 and
they can be constructed as crossed products with  the
dual of ${\cal G}$. Decisive
for this construction is the property that the relative commutant of the fixed
point algebra ${\cal A}$ 
under the ${\cal G}$ action 
is trivial and this 
requires 
 the triviality of
the center of ${\cal A}$ (see [DR1]).

Later it turned out that Hilbert systems with nontrivial center
 ${\cal Z}$
of ${\cal A}$
may be also of interest. For example, if 
${\cal G}=T$
then there is an intrinsic link between the circle group $T$ and 
the group of all
continuous $T$-valued functions on spec ${\cal Z}$, 
denoted  by  
$C(\mbox{spec}\,{\cal Z}\rightarrow T)$, given by its isomorphy to
$\mbox{stab}\,{\cal A}$.
This is due to the fact that the property of the model 
to be Galois closed breaks down (see [BL1]).

If $\mbox{spec}\,{\cal Z}$ can be considered as a "base space" and 
$T$ as a "global gauge group" then this means that
$\mbox{stab}\,{\cal A}$ appears as a  "local
gauge group".

We adopt 
here a pure mathematical point of view. 
Nevertheless, the property just mentioned suggests to construct
models such that these interpretations may be justified.

We concede that only the last example of the present paper
points into this direction. The interpretation of ${\cal Z}$ in
the other examples is rather that of a space of labels for
multiplicity resp. degeneracy.

Note that in the case of abelian groups ${\cal
G}$ the extension problem is solved in general terms (see,
for example, [B1],[BL2]). Moreover, the case ${\cal G}=T$ 
is of
special interest because of the special property of the stabilizer
$\mbox{stab}\,{\cal A}$
in this case, mentioned above.

This paper presents constructions for the action 
of $T$ 
 which are closely related to
fermion quantization in general,
 and in particular to the implementation on the corresponding Fock spaces
of the automorphisms of the CAR algebra defined by 
loop groups. Techniques and ideas from [CR1] and
other related papers are used and partly simplified, for example by 
use of the selfdual framework.

First a construction is given on the reference
Hilbert space $\cal H$ 
without use of quantization procedures which in
turn yields a further example with nonnegative spectrum by 
quantising in the Fock space using the ``positive energy'' criterion.
 The third construction is a general procedure within the selfdual framework of
fermion quantization (see [A1]). This procedure is then applied to an example
where the implementation of the automorphisms defined by 
loop groups is used. In this example 
the center is calculated explicitly. 
 According to the general observation
mentioned above this means simply that in this model the group
$C(\mbox{spec}\,{\cal Z}\rightarrow T)$ 
is nothing 
other 
 than the stabilizer of the fixed point
algebra ${\cal A}$.

Finally we present an example, using a simple tensor product 
construction, where
$\mbox{spec}\,{\cal Z}=S^{1}$
is satisfied. This example shows that, at least in special cases, the
interpretation of
$\mbox{stab}\,{\cal A}$
as a "local gauge group" seems to be reasonable.

\vspace{0.3cm}

DEFINITION. A Hilbert system for a 
compact abelian group ${\cal G}$ is a C*-system
 $\{{\cal F},\alpha_{\cal G}\}, \alpha_{\cal G}\subset \mbox{aut}\,{\cal F},
\,\alpha_{\cal G}$ continuous w.r.t. the topology of 
 pointwise norm convergence, with spec$\,\alpha_{\cal
G}=\hat{\cal G}$ such that each spectral subspace
 contains a unitary.

As usual, the stabilizer
$\mbox{stab}\,{\cal A}$
for a unital $C^{\ast}$-subalgebra
${\cal A}\subset {\cal F}$
is defined by
$\mbox{stab}\,{\cal A}:=\{\beta\in\mbox{aut}\,{\cal F}: \beta(A)=A\,\mbox{for
all}\,A\in {\cal A}\}.$

\vspace{0.3cm}
 
The
terminology 
 Hilbert system can be traced back to Doplicher and Roberts [DR2]
where - in the case just mentioned -
 the spectrum of $\alpha_{\cal G}$ is called
the Hilbert spectrum.

\section{Selected properties of Hilbert $C^{\ast}$-systems}

For convenience we recall some properties of Hilbert systems for the action T
which are useful for the following constructions (cf. [BL1]).

Let
$\{{\cal F},\alpha_{T}\}$
be a Hilbert system,
$\Pi_{n},\,n\in Z$
its spectral projections and
${\cal F}_{n}:=\Pi_{n}{\cal F}$
the corresponding spectral subspaces.
${\cal F}_{0}=:{\cal A}$
is the fixed point algebra and
${\cal Z}$
its center.

If
$V\in{\cal F}_{1}$
is a unitary then
$V^{n}\in {\cal F}_{n},\,{\cal F}_{n}={\cal A}V^{n}$
and
$${\cal F}=\mbox{clo}_{\Vert\cdot\Vert}\{\sum_{n\in Z}A_{n}V^{n},\,
A_{n}\in {\cal A}\;\mbox{finite sum}\}. $$
Put
$\kappa :=\mbox{Ad}\,V|{\cal A}$
then
$\kappa\in\mbox{aut}\,{\cal A}.$
If
$\tilde{V}:=UV$
with
$U\in{\cal U}({\cal A})$
then
$\mbox{Ad}\,\tilde{V}|{\cal A}=\mbox{Ad}\,U\circ \kappa =:\tilde{\kappa},$
i.e.
$\kappa$ and $\tilde{\kappa}$ are unitarily equivalent. If
$\kappa=\tilde{\kappa}$,
i.e.
$UVAV^{-1}U^{-1}=VAV^{-1}$
for all
$A\in {\cal A}$
then
$UBU^{-1}=B$
for all
$B\in {\cal A}$
and
$U\in {\cal Z}$
follows. This means:
$\kappa$
determines uniquely only
$V{\cal Z}={\cal Z}V$.
The latter equation is true because of
$\kappa({\cal Z})={\cal Z}.$
For
$\alpha,\beta\in\mbox{aut}\,{\cal A}$
the space of intertwiners is defined by
\[
(\alpha,\beta):=\{X\in{\cal A}:X\alpha(A)=\beta(A)X\,\mbox{for all}\,
A\in {\cal A}\}.
\]
Then one obtains: 
$(\kappa^{n},\iota)=\{0\}$
for all
$n\neq 0$
or
$"\kappa^{n}$
and
$\kappa^{m}$
are mutually disjoint for
$m\neq n$"
iff
${\cal A}'\cap {\cal F}={\cal Z}$
(see [BL1]. In this case the stabilizer turns out to be isomorphic to
${\cal U}({\cal Z})$
where the isomorphism is given by
\[
\mbox{stab}\,{\cal A}\ni\beta\rightarrow Z(\beta)\in{\cal U}({\cal Z}):
\beta(V)=VZ(\beta).
\]
Note that
${\cal U}({\cal Z})\cong C(\mbox{spec}\,{\cal Z}\rightarrow T).$
Moreover, each element
$\beta\in\mbox{stab}\,{\cal A}$
commutes with the starting group
$\alpha_{T}$,
i.e.
$\beta\circ\alpha_{\zeta}=\alpha_{\zeta}\circ\beta$
for all
$\zeta\in T$
and, if
$V\in{\cal F}_{1}$
then
$\beta(V)\in{\cal F}_{1}$
and
$\mbox{Ad}\,\beta(V)|{\cal A}=\kappa,$
i.e. the automorphism of 
${\cal A}$
generated by $V$ is invariant w.r.t. application of $\beta$. These properties of
$\mbox{stab}\,{\cal A}$
point to possible interpretation in suitable models.

Concerning the center
${\cal Z}$
note the following observation (see [BL1]): If
$\omega_{0}$
is a state of
${\cal A}$
such that the corresponding GNS-representation
$\pi_{0}$
is faithful and
$\pi$
is the GNS-representation of
${\cal F}$
w.r.t.
$\omega(F):=\omega_{0}(\Pi_{\iota}f)$
and
$U(T)$
the (unique) implementer of
$\alpha_{T}$
then
$\pi({\cal A})\cap U(T)''=C1$
and
$\pi({\cal A})''\subseteq U(T)'$
follow. Now, if
${\cal Z}\supset C1$
then
\[
\pi({\cal A})''\subset U(T)'
\]
follows, i.e.the breakdown of the Galois closedness implies the "gap" described
by this proper inclusion (in the case
${\cal Z}=C1$
one has equality). This means the group $U(T)$ does not determine
$\pi({\cal A})$
completely. On the other hand the representation
$\pi_{0}$
is not irreducible which points to a "degeneracy of the vacuum" in corresponding
models.

\section{Constructions using the regular representation of $T$ and the bilateral shift} 
\subsection{A direct approach}

The unit circle is denoted by $S^{1}$ (as a topological space) and 
by $T$ (as the 1-torus
group).
 By $e_{n},n\in Z,$ we denote the canonical orthonormal basis $e_{n}(\xi):=
\xi^{n}$ in $L^{2}(S^{1})$ and by $P_{n}:=(e_{n},\cdot)e_{n}$ the corresponding
1-dimensional projection. Let ${\cal K}$ be a Hilbert space. We put
${\cal H}_{0}:=L^{2}(S^{1}, {\cal K}).$ 
The regular representation of $T$ on ${\cal H}$ is denoted by
$U_{\zeta},\,\zeta\in T$, 
 where 
$$(U_{\zeta}f)(\xi):=f(\zeta\xi),\,f\in {\cal H}_{0}.$$ 
Putting $E_{n}:=P_{n}\otimes 1_{\cal K}$ we have
$U_{\zeta}=\sum_{n\in Z}\zeta^{n}E_{n},$
i.e. $E_{n}$ is the isotypical projection w.r.t. the label 
$n$, 
and  with $U_T=\{U_\zeta | \zeta\in T\}$ 
we have spec$\,U_{T}=Z.$ The
bilateral shift on $L^{2}(S^{1})$ w.r.t. the canonical basis is
 denoted by $V_{0}:
V_{0}e_{n}:=e_{n+1},n\in Z.$ Then
$V:=V_{0}\otimes 1_{\cal K}$ satisfies

\begin{equation}
U_{\zeta}VU_{\zeta}^{-1}=\zeta V, \quad \zeta \in T,
\end{equation}
because
$$U_{\zeta}VU_{\zeta}^{-1}e_{n}\otimes k=\zeta^{-n}U_{\zeta}Ve_{n}\otimes k=
\zeta^{-n}U_{\zeta}e_{n+1}\otimes k=\zeta^{-n}\zeta^{n+1}e_{n+1}\otimes k=
\zeta Ve_{n}\otimes k.$$
Using $U_{T}$ and $V$ one can construct Hilbert systems w.r.t. $T$
 as follows.
Let $C^{*}(U_{T},V)$ be the C*-algebra generated by $U_T,V$. 

\vspace{0.3cm}

\noindent PROPOSITION 1. 
{\em Let} ${\cal F}$ {\em be a} $C^{*}-${\em algebra with}
\begin{equation}
C^{*}(U_{T},V)\subseteq {\cal F}\subseteq {\cal L}({\cal H}_{0}).
\end{equation}
{\em Then}:\\
 (I) $\alpha_{\zeta}:=\mbox{Ad}\,U_{\zeta}|{\cal F}\in
\mbox{aut}\,{\cal F},
\,\zeta\in T ,$ 
{\em and} $\zeta\rightarrow\alpha_{\zeta}$ {\em is 
continuous w.r.t.  pointwise norm
convergence.}\\
 (II) $\{{\cal F},\alpha_{T}\}$ {\em is a 
Hilbert system and the spectral
subspaces} ${\cal F}_{n}$ {\em are given by}
\[
{\cal F}_{n}={\cal A}V^{n},\quad n\in Z,
\]
{\em where} ${\cal A}$ {\em denotes the fixed point algebra}
${\cal A}:={\cal F}_{0}=U_{T}'\cap {\cal F}.$\\
(III) {\em The center} ${\cal Z}$ {\em of} ${\cal A}$ {\em satisfies}
\begin{equation}
{\cal Z}={\cal A}'\cap {\cal F},
\end{equation}
{\em i.e. it coincides with the relative commutant of} ${\cal A}$. {\em Moreover}
$C^{*}(U_{T})\subseteq {\cal Z}.$

\vspace{0.3cm}

\noindent {\bf Proof}.
 These facts are elementary consequences of the definitions. 
 Note that
${\cal Z}={\cal A}\cap{\cal A}'={\cal F}\cap U_{T}'\cap{\cal A}'=
{\cal F}\cap (U_{T}\cup {\cal A})'={\cal F}\cap {\cal A}'$
because $U_{T}\subset U_{T}',$ hence $U_{T}\subset {\cal A}$ follows. 
Note further
$ {\cal A}\subseteq U_{T}'$ hence 
$U_{T}\subset U_{T}''\subseteq {\cal A}'$ and $U_{T}\subset {\cal Z}$
follows. $\Box$

\vspace{0.3cm}

In particular, we consider the extremal special 
cases of the inclusion (2). First let
${\cal F}:={\cal L}({\cal H}_{0})$.
Then ${\cal A}=U_{T}'$ , i.e.
$A\in {\cal A}$ iff $AE_{n}=E_{n}A$ for all $n\in Z$
which means that there is a one-to-one correspondence
$A \leftrightarrow \{B_{n}\}_{n\in Z},\,B_{n}\in {\cal L}({\cal K}),\,
\mbox{sup}_{n}\,\Vert B_{n}\Vert < \infty,$
i.e. ${\cal A}$ corresponds to the so-called 
algebra of all diagonalizable operators.
Moreover,
${\cal A}'=U_{T}''$
and
${\cal Z}=U_{T}'\cap U_{T}''=\{\sum_{n\in Z}\alpha_{n}E_{n},\,
\mbox{sup}_{n}\,\vert \alpha_{n}\vert < \infty\},$
i.e. ${\cal Z}$ is isomorphic to the algebra of all bounded 
complex-valued functions on
$Z$. Then $\mbox{spec}\,{\cal Z}$ coincides with the 
Stone-Cech-compactification of $Z$.
Second let
${\cal F}:=C^{\ast}(U_{T},V).$ Then
${\cal F}=\mbox{clo}_{\Vert\cdot\Vert}\{\sum_{n\in Z}C_{n}V^{n},\,
\mbox{finite sum}\,,
C_{n}\, \mbox{polynomial w.r.t.}\, U_{T}\}$
because $U_{\zeta}V=\zeta VU_{\zeta}.$ Hence
${\cal A}=C^{\ast}(U_{T})$
follows, i.e. ${\cal A}={\cal Z}$ .

\subsection{Quantized version with nonnegative spectrum}

Applying quasifree 
fermion quantization to the foregoing example one gets Hilbert systems
where the representation (of $T$),
 which defines via ``Ad" 
 the automorphism group,
has nonnegative spectrum. Recall that the generator $H$ of
$U_{e^{it}}=e^{itH},\,H=\sum_{n}E_{n}$
may be used to define by implementation, the free
fermion Hamiltonian or, in fermionic conformal field theory models,
a generator of the Virasoro algebra. 
 In the
following we assume 
$\dim\,{\cal K}<\infty.$
We put
\[
E_{\geq 0}:=\sum_{n\geq 0}E_{n},\quad E_{<0}:=1-E_{\geq 0}.
\]
Now let $\gamma$ be an arbitrary conjugation on ${\cal H}_{0}$, i.e.
$\gamma$ is anti-unitary and
$\gamma^{2}=1.$
Put ${\cal H}:={\cal H}_{0}\oplus {\cal H}_{0}.$
Then
$\Gamma$
defined by
$\Gamma (f,g):=(\gamma g,\gamma f)$
is a conjugation on ${\cal H}$. Adopting the terminology of the selfdual
framework for the fermion algebra
$CAR({\cal H},\Gamma)$ all operators 
$\phi(U)$
defined by
\begin{equation}
\phi(U)(f,g):=(Uf,\gamma U\gamma f),\quad 
(f,g)\in{\cal H},\quad U\in {\cal U}({\cal H}_{0}),
\end{equation}
are Bogoljubov unitaries on ${\cal H}$ defining Bogoljubov automorphisms
$\alpha_{U}$ of $CAR({\cal H},\Gamma)$ and the projection
$\Pi$ defined by
\[
\Pi(f,g):=(E_{\geq 0}f,\gamma E_{<0}\gamma g),\quad (f,g)\in{\cal H},
\]
is a basis projection, i.e. it defines a Fock representation
$\pi$
of
$CAR({\cal H},\Gamma)$
on the Fock space
${\cal F}_{\Pi}({\cal H},\Gamma).$
According to the well-known implementation criterion the operators (4) are
implementable by unitaries
$\Phi(U)$
on the Fock space w.r.t. $\Pi$, this means
\[
\pi\circ\alpha_{U}=\mbox{Ad}\,\Phi(U)\circ\pi,
\]
iff
$E_{\geq 0}UE_{<0}$ and $E_{<0}UE_{\geq 0}$
are Hilbert-Schmidt operators, i.e. are from
${\cal L}_{2}({\cal H}_{0})\,$
(note that these are two independent conditions). The implementer
$\Phi(U)$ is unique $\mbox{mod}\,T1$ 
 in general.

Note that $\phi(V)$ is implementable. Choose, for example,
$E_{\geq 0}VE_{<0}.$
The condition
$E_{\geq}VE_{<0}\in {\cal L}_{2}({\cal H}_{0})$
is equivalent to
$VE_{\geq 0}-E_{\geq 0}V\in {\cal L}_{2}({\cal H}_{0}).$
That is, according to
$VE_{n}=E_{n+1}VE_{n}$
or
$V=\sum_{n}E_{n+1}VE_{n}$
we obtain
\[
\sum_{n\geq 0}E_{n}\sum_{m}E_{m+1}VE_{m}-\sum_{m}E_{m+1}VE_{m}
\sum_{n\geq 0}E_{n}=
\]
\[
\sum_{m\geq -1}E_{m+1}VE_{m}-\sum_{m\geq 0}E_{m+1}VE_{m}=E_{0}VE_{-1}
\]
which is finite-dimensional.

Furthermore, 
$\phi(U_{\zeta})$
is implementable without phase ambiguity because it commutes with
$\Pi$.
Then we can state

\vspace{0.3cm}

\noindent LEMMA 1. {\em The implemented strongly continuous unitary group}
$\Phi(U_{T})$
{\em on}
${\cal F}_{\Pi}({\cal H},\Gamma)$
{\em has nonnegative spectrum,}
$\mbox{spec}\,\Phi(U_{T})\geq 0.$

\vspace{0.3cm}

\noindent Proof. One has
$\Pi{\cal H}=E_{\geq 0}{\cal H}_{0}\oplus\gamma E_{<0}\gamma{\cal H}_{0}.$
Obviously,
$\mbox{spec}\,U_{T}|E_{\geq 0}{\cal H}_{0}\geq 0.$
But also
$\mbox{spec}\,\gamma U_{T}\gamma|\gamma E_{<0}\gamma{\cal H}_{0}\geq 0,$
because
$\gamma U_{\zeta}\gamma =\sum_{n}\zeta^{-n}\gamma E_{n}\gamma$
and a label $n<0$ gives a positive spectral value $-n.\,\Box$

\vspace{0.3cm}

The relation (1) implies the corresponding relation for the implementers:

\begin{equation}
\Phi(U_{\zeta})\Phi(V)\Phi(U_{\zeta})^{-1}=\zeta\Phi(V),\quad \zeta\in T.
\end{equation}
This leads to the following Hilbert systems:

\vspace{0.3cm}

\noindent PROPOSITION 2. {\em Let}
${\cal F}$
{\em be a} $C^{\ast}$-{\em algebra on the Fock space}
${\cal F}_{\Pi}({\cal H},\Gamma)$
{\em with}
\[
C^{\ast}(\Phi(U_{T}),\Phi(V))\subseteq {\cal F}\subseteq
C^{\ast}(\Phi(U),\,U\in{\cal U}({\cal H})\,,\phi(U)\,\mbox{implementable}).
\]
{\em Then:}\\
(I)$\,\alpha_{\zeta}:=\mbox{Ad}\,\Phi(U_{\zeta})|{\cal F}
\in\mbox{aut}\,{\cal F},\,\zeta\in T,$
{\em and}
$\zeta\rightarrow\alpha_{\zeta}$
{\em is continuous w.r.t. the pointwise norm convergence.}\\
(II) $\{{\cal F},\alpha_{T}\}$
{\em is a Hilbert system and the spectral subspaces}
${\cal F}_{n},\,n\in Z,$
{\em are given by}
\[
{\cal F}_{n}={\cal A}\Phi(V)^{n},\quad n\in Z,
\]
{\em where}
${\cal A}$
{\em denotes the fixed point algebra}
${\cal A}:={\cal F}_{0}={\cal F}\cap \Phi(U_{T})'.$\\
(iii){\em The center}
${\cal Z}$
{\em of}
${\cal A}$
{\em satisfies}
\[
{\cal Z}={\cal A}'\cap {\cal F}
\]
{\em and one has}
\[
C^{\ast}(\Phi(U_{T}))\subseteq {\cal Z}.
\]

\vspace{0.3cm}

\noindent {\bf Proof}. Similar to Proposition 1. $\Box$

\vspace{0.3cm}

According to this result the question arises: 
 find conditions for proper
extensions
${\cal F}\supset C^{\ast}(\Phi(U_{T}),\Phi(V))$
such that
${\cal Z}=C^{\ast}(\Phi(U_{T})).$

\section{Constructions on the quantized level for the identical representation
of $T$}
\subsection{A general procedure}

The aim of this section is to present an 
approach for the construction of Hilbert
systems for the action of $T$ 
such that the center of 
the fixed point algebra can be calculated explicitly.

The starting point for this approach is the selfdual framework of the fermion
algebra
$CAR({\cal H},\Gamma)$
and the essential tools are Fock representations, implementation theorems and
the CAR-CCR correspondence given by the so-called Schwinger term.

\subsubsection{Preliminary remarks}

We start with an infinite-dimensional Hilbert space
${\cal H}_{0}$
and an arbitrary conjugation
$\gamma$
on
${\cal H}_{0}$.
We put
${\cal H}:={\cal H}_{0}\oplus {\cal H}_{0}$
and define the conjugation
$\Gamma$ on ${\cal H}$ as before by
$\Gamma(f,g):=(\gamma g,\gamma f),\,(f,g)\in {\cal H}.$
For unitaries
$U\in {\cal U}({\cal H}_{0})$
the unitary
$\phi(U)$ defined by

\begin{equation}
\phi(U)(f,g):=(Uf,\gamma U\gamma g),\quad (f,g)\in {\cal H}
\end{equation}
is a Bogoljubov unitary w.r.t.
$\Gamma$ on ${\cal H}$.
Further we choose an arbitrary orthoprojection $P$ on 
${\cal H}_{0},\,0<P<1,$
with
$\dim P=\dim P^{\bot}=\infty,$
where
$P^{\bot}:=1-P.$
Then the projection $\Pi$ on ${\cal H}$ defined by

\begin{equation}
\Pi(f,g):=(Pf,\gamma P^{\bot}\gamma g),\quad (f,g)\in {\cal H}
\end{equation}
is a basis projection w.r.t.
$\Gamma$ and ${\cal H}$. To $\Pi$ there corresponds a unique Fock representation
$\pi$ of
$CAR({\cal H},\Gamma)$
on the corresponding Fock space, denoted by
${\cal F}_{\Pi}({\cal H},\Gamma).$
Bogoljubov unitaries $V$ on ${\cal H}$ define Bogoljubov automorphisms
$\alpha_{V}\in \mbox{aut}\,CAR({\cal H},\Gamma).$

The question arises: under which conditions can 
$\alpha_{V}$ be implemented w.r.t.
$\pi$ by an Ad-automorphism 
 on the Fock space? 
That is, we need the 
existence of a unitary $\Phi(V)$ satisfying
\[
\pi\circ \alpha_{V}=\mbox{Ad}\,\Phi(V)\circ \pi.
\]
This 
is answered by the well-known 
implementation criterion (see, for example, [A1]).
In general,
$\Phi(V)$
is unique
$\mbox{mod}\,T1$ and $\Phi(V)$
is called an implementer of $V$ on the Fock space. The implementation criterion
for the special Bogoljubov unitaries 
$\phi(U)$
says: $\phi(U)$ has an implementer
$\Phi(U):=\Phi(\phi(U))$ iff
$PUP^{\bot}$ and $P^{\bot}UP$
are Hilbert-Schmidt, i.e. members of
${\cal L}_{2}({\cal H}_{0}).$
Since
\[
\Pi\phi(U)\Pi = \mbox{diag}\,\{PUP,\gamma P^{\bot}UP^{\bot}\gamma\}
\]
one obtains 
\[
0=\mbox{ind}\,\Pi\phi(U)\Pi |\Pi{\cal H}=\mbox{ind}\,PUP|P{\cal H}_{0}+
\mbox{ind}\,P^{\bot}UP^{\bot}|P^{\bot}{\cal H}_{0}.
\]

Denote by 
${\cal U}_{res}\subset {\cal U}({\cal H}_{0})$
the subgroup of all implementable unitaries on
${\cal H}_{0}.$
In particular, the scalar unitaries
$U_{\lambda}:=\lambda 1,\,\lambda\in T,$
are implementable, even without phase ambiguity because
$\phi(U_{\lambda})$
commutes with 
$\Pi$. 
The corresponding unique implementer is denoted by
$\Phi(\lambda).$
Thus the `identical representation'
$T\ni\lambda\rightarrow \lambda 1$
on
${\cal H}_{0}$
is implemented on the Fock space by
$\lambda\rightarrow\Phi(\lambda)\in{\cal U}({\cal F}_{\Pi}({\cal H},\Gamma))$
whereas
${\cal U}_{res}\ni U\rightarrow\Phi(U)$
is a projective unitary representation.

To each
$U\in{\cal U}_{res}$
there corresponds uniquely an {\em index}
$q(U)\in Z$,
defined by

\begin{equation}
\Phi(\lambda)\Phi(U)\Phi(\lambda)^{-1}=\lambda^{q(U)}\Phi(U),\quad \lambda\in T.
\end{equation}
According to a result of Carey, Hurst, O'Brien [CHO1]
$q(U)$
is given by
\[
q(U)=\mbox{ind}\,PUP|P{\cal H}_{0}.
\]
This implies that there are unitaries
$V_{1}\in{\cal U}_{res}$
satisfying
$q(V_{1})=1.$
Then
$q(V_{1}^{n})=n,\,n\in Z,$
follows. Note that
$q(U_{1}U_{2})=q(U_{1})+q(U_{2})$ 
for
$U_{1},U_{2}\in{\cal U}_{res}.$
The set of all
$U\in {\cal U}_{res}$
with
$q(U)=n$
is denoted by
${\cal U}_{res}({n})$.
Thus
${\cal U}_{res}({n})\neq \emptyset.$
If
$V_{1}\in {\cal U}_{res}({n})$
then
\[
{\cal U}_{res}({n})=\{UV_{1}:U\in {\cal U}_{res}(0)\}
=\{V_{1}U:U\in{\cal U}_{res}({0})\}. 
\]
Putting
$\alpha_{\lambda}:=\mbox{Ad}\,\Phi(\lambda)|C^{\ast}(\Phi({\cal U}_{res}))$
we obtain a "maximal" Hilbert system 
$\{C^{\ast}(\Phi({\cal U}_{res})),\alpha_{T}\}$
where
${\cal A}=C^{\ast}(\Phi({\cal U}_{res}(0))).$
Note that
$$V_{1}UV_{1}^{-1},V_{1}^{-1}UV_{1}\in {\cal U}_{res}({0})$$
if
$U\in{\cal U}_{res}({0}), V_{1}\in{\cal U}_{res}({1}),$
i.e.
$$\mbox{Ad}\,V_{1}|C^{\ast}(\Phi({\cal U}_{res}(0)))\in\mbox{aut}
\,C^{\ast}(\Phi({\cal U}_{res}({0}))).$$
Let
$V_{1}\in{\cal U}_{res}({1})$
be fixed. We put
$${\cal F}_{V_{1}}:=C^{\ast}(\Phi(T),\Phi(V_{1})).$$
Then
$\{{\cal F}_{V_{1}},\alpha_{T}|{\cal F}_{V_{1}}\}$
is a "minimal" Hilbert system, because from (8) we obtain
$$\Phi(V_{1})\Phi(\lambda)\Phi(V_{1})^{-1}=\lambda^{-1}\Phi(\lambda)$$
and this means that
$\mbox{Ad}\,V_{1}|C^{\ast}(\Phi(T))\in\mbox{aut}\,C^{\ast}(\Phi(T))$
which implies that
${\cal F}_{V_{1}}\cap\Phi(T)'=C^{\ast}(\Phi(T)).$
Therefore we obtain a preliminary result which corresponds to Proposition 2:

\vspace{0.3cm}

\noindent PROPOSITION 3. {\em Let}
${\cal F}$
{\em be a}
$C^{\ast}$-{\em algebra on the Fock space}
${\cal F}_{\Pi}({\cal H},\Gamma)$
{\em with}
\[
C^{\ast}(\Phi(T),\Phi(V_{1}))\subseteq {\cal F}\subseteq C^{\ast}
(\Phi({\cal U}_{res})).
\]
{\em Then:}\\
 (I)
$\alpha_{\lambda}:=\mbox{Ad}\,\Phi(\lambda)|{\cal F}\in\mbox{aut}\,{\cal F},\,
\lambda\in T,$
{\em and}
$\lambda\rightarrow \alpha_{\lambda}$
{\em is continuous w.r.t. pointwise norm convergence.}\\
 (II)
$\{{\cal F},\alpha_{T}\}$
{\em is Hilbert and the spectral subspaces}
${\cal F}_{n}$
{\em are given by}
\[
{\cal F}_{n}={\cal A}\Phi(V_{1}^{n}),\quad n\in Z,
\]
{\em where}
${\cal A}$
{\em denotes the fixed point algebra}
\[
{\cal A}:={\cal F}_{0}={\cal F}\cap\Phi(T)'\subseteq C^{\ast}(\Phi(
{\cal U}_{res}({0}))).
\]
(III)
{\em The center}
${\cal Z}$
{\em of}
${\cal A}$
{\em satisfies}
\[
{\cal Z}={\cal A}'\cap {\cal F}
\]
{\em and}
\begin{equation}
C^{\ast}(\Phi(T))\subseteq {\cal Z}.
\end{equation}

\vspace{0.3cm}

\noindent {\bf Proof}. Immediate from the preceeding discussion. $\Box$

\vspace{0.3cm}

If
${\cal F}$
is the "minimal" Hilbert system mentioned above then
${\cal A}=C^{\ast}(\Phi(T))$
is abelian and we have equality in (9). The problem is to construct proper
intermediate Hilbert systems
$\{{\cal F},\alpha_{T}\}$
such that
${\cal Z}=C^{\ast}(\Phi(T)).$
In the following we present an abstract approach to produce such Hilbert
systems.

\subsubsection{The Schwinger term}

According to 3.1.1 we have
$T1_{{\cal H}_{0}}\subset{\cal U}_{res}({0}).$
Now we consider subgroups
${\cal V}_{0}\subset{\cal U}_{res}({0})$ 
with
$T1_{{\cal H}_{0}}\cap{\cal V}_{0}=\{1_{{\cal H}_{0}}\}$
and which are generated by 
 their 1-parameter subgroups
$e^{itA}$
where the generators $A$ satisfy
$A=A^{\ast}\in {\cal L}({\cal H}_{0})$
and
$\gamma A\gamma =A.$
Then
$\phi(e^{itA})=e^{it\phi(A)}$
where 
$\phi(A)$
is defined by
\[
\phi(A)(f,g):=(Af,-Ag),\,(f,g)\in{\cal H}.
\]
$\phi(A)$
satisfies
$\Gamma\phi(A)\Gamma=-\phi(A).$
By ${\cal L}$ we denote the real-linear subspace of all these generators. In
what follows we refer to Araki [A1] and quote some results, for convenience. The
{\em first requirement} for ${\cal L}$ is
$\Pi\phi(A)\Pi^{\bot}\in {\cal L}_{2}({\cal H})$
for
$A\in {\cal L},$
which means
$PAP^{\bot},P^{\bot}AP\in{\cal L}_{2}({\cal H}_{0}).$
Then, according to [A1,Theorem 6.10,p.78] one has the following result:

The group
$e^{itA}$
has a special continuity property (P-norm continuity) and the implementation
$\Phi(e^{itA})$
is a unitary strongly continuous group
\[
\Phi(e^{itA})=e^{it\Phi(\phi(A))},\,\Phi(\phi(A))\,\mbox{selfadjoint on}
\,{\cal F}_{\Pi}({\cal H},\Gamma),\,A\in{\cal L},
\]
(where we use for
$\phi(A)$
the same implementation symbol as for unitaries). The set of all generators
$\Phi(\phi(A))$
has a common dense domain and on this domain the commutator
$[\Phi(\phi(A_{2})),\Phi(\phi(A_{1}))]$
can be calculated:
\[
[\Phi(\phi(A_{2})),\Phi(\phi(A_{1}))]=i\Phi([\phi(A_{2}),\phi(A_{1})])+
\]
\[
\frac{1}{2}\mbox{tr}\{\Pi\phi(A_{1})\Pi^{\bot}\phi(A_{2})\Pi-
\Pi\phi(A_{2})\Pi^{\bot}\phi(A_{1})\Pi\}=
\]
\[
i\{\Phi(\phi([A_{2},A_{1}]))+\mbox{Im\,tr}(PA_{1}P^{\bot}A_{2}P)\}
\]
where the second term is known as the Schwinger term. We put
\[
s(A_{1},A_{2}):=2\mbox{Im\,tr}\,(PA_{1}P^{\bot}A_{2}P),\quad A_{1},A_{2}
\in {\cal L}.
\]
$s(\cdot,\cdot)$
is symplectic bilinear on
${\cal L}$.
Further we put
\[
\langle A_{1},A_{2}\rangle := \mbox{tr}\,(PA_{1}P^{\bot}A_{2}P),\quad A_{1},
A_{2}\in {\cal L}.
\]
Then
$\langle\cdot,\cdot\rangle$
is a semi-scalar product on
${\cal L}$
and
$\langle A,A\rangle =0$
iff
$PAP^{\bot}=P^{\bot}AP=0.$

The {\em second requirement} is that
${\cal V}_{0}$ 
is abelian, or equivalently, that the generators
$A\in{\cal L}$
all commute.
 Then
\[
[\Phi(\phi(A_{2})),\Phi(\phi(A_{1}))]=i\mbox{Im\,tr}(PA_{1}P^{\bot}
A_{2}P),\quad A_{1},A_{2}\in {\cal L}.
\]
The {\em third requirement} is:
$\langle\cdot,\cdot\rangle$
is a scalar product on
${\cal L}$.
This implies that $s$  is non-degenerate 
 on
${\cal L}$.
Let
${\cal W}:=CCR({\cal L},s)$
be the Weyl algebra for the parameters
$\{{\cal L},s\}.$
Then, according to [A1,Theorem 7.1, p.121] or [CR1] 
 one has:

The representation $\rho$ of ${\cal W}$ defined by
\[
\rho(W(A)):=\Phi(e^{iA})=e^{i\Phi(\phi(A))},\quad A\in {\cal L},
\]
on the Fock space
${\cal F}_{\Pi}({\cal H},\Gamma)$
is the Fock representation w.r.t. the generating functional
\[
f(A):=e^{-\frac{1}{4}\langle A,A \rangle},
\]
where $W(A)$ denotes the abstract Weyl generators for
${\cal W}$.

Note that 
$\rho({\cal W})$
is simple and therefore it has trivial center. Now recall that
$T1_{{\cal H}_{0}}\cdot{\cal V}_{0}$ 
is a direct product. This gives
\[
{\cal A}:=C^{\ast}(\Phi(T),\rho({\cal W}))\cong 
C^{\ast}(\Phi(T))\otimes \rho({\cal W})
\]
and
\[
{\cal Z}={\cal Z}({\cal A})\cong 
{\cal Z}(C^{\ast}(\Phi(T)))\otimes {\cal Z}(\rho({\cal W}))\cong
C^{\ast}(\Phi(T)).
\]
The {\em last requirement} refers to some
$V_{1}\in{\cal U}_{res}({1})$ 
and its 
 connection to 
${\cal L}.$
It says that
\[
V_{1}{\cal L}V_{1}^{-1}={\cal L},
\]
i.e.
$\mbox{Ad}\,V_{1}$ 
acts on
${\cal L}$
as an automorphism. This implies
\[
\Phi(V_{1})\rho({\cal W})\Phi(V_{1})^{-1}=\rho({\cal W}),
\]
i.e.
$\mbox{Ad}\,\Phi(V_{1})|\rho({\cal W})\in\mbox{aut}\,\rho({\cal W})$ 
hence
\begin{equation}
\mbox{Ad}\,\Phi(V_{1})|{\cal A}\in\mbox{aut}\,{\cal A}
\end{equation}
follows and we obtain

\vspace{0.3cm}

\noindent PROPOSITION 4. {\em Let}
${\cal F}:=C^{\ast}({\cal A},V_{1})\in{\cal L}({\cal F}_{\Pi}({\cal H},
\Gamma))$
{\em and}
$\alpha_{\lambda}:=\mbox{Ad}\,\Phi(\lambda)|{\cal F},\,\lambda\in T.$ 
{\em Then}
$\alpha_{T}\subset \mbox{aut}\,{\cal F}$
{\em and}
$\{{\cal F},\alpha_{T}\}$
{\em is a  Hilbert system 
where the fixed point algebra coincides with}
${\cal A}.$
{\em The center}
${\cal Z}$
{\em of}
${\cal A}$
{\em satisfies}
\[
{\cal Z}={\cal A}'\cap {\cal F}
\]
{\em and}
\[
{\cal Z}= C^{\ast}(\Phi(T)).
\]

\vspace{0.3cm}

\noindent{\bf Proof}. This follows immediately from the discussion
preceeding the statement of the proposition, in particular (10). $\Box$

\subsection{An example using the implementation of the loop group}

In this section we present an example where all requirements of the foregoing
procedure are satisfied. Put
\[
{\cal H}_{0}:=L^{2}(S^{1}),
\]
\[
\gamma: (\gamma (x))(\zeta):=\overline{x(\zeta)},\quad \mbox{complex 
conjugation,}
\]
\[
P:=P_{\geq 0},
\]
where
$P_{\geq 0}$
denotes the projection onto the ``nonnegative terms" of the regular
representation of $T$ on
$L^{2}(S^{1})$,
i.e.
$P_{\geq 0}=\sum_{n\geq 0}P_{n},$
where the
$P_{n}$
are the 1-dimensional projections of Section 2. Let
${\cal U}_{res}$
be as before. In this example first we choose a subgroup
$\tilde{\cal V}_{0}\subset{\cal U}_{res}({0})$
which is defined as follows: Let
$L(T):=C^{\infty}(S^{1}\rightarrow T)$
be the {\em loop group} of $T$. 
 The functions
$f\in L(T)$
are interpreted as {\em multiplication operators}
\[
f\rightarrow U_{f}: (U_{f}(x))(\zeta):=f(\zeta)x(\zeta),\quad x\in
L^{2}(S^{1}).
\]
It is well known that
$U_{L(T)}\subset {\cal U}_{res}$
(see,for example, [PS1]). Further we have
\[
\mbox{ind}\,PU_{f}P|P{\cal H}_{0}=w(f)
\]
where $w(f)$ is the {\em winding number} of $f$ (see [CH1]), i.e. we have
$q(U_{f})=w(f).$
Now we put
\[
\tilde{\cal V}_{0}:=\{U_{f}\in U_{L(T)}: w(f)=0\}.
\]
Obviously
$w(f)=0$
implies
$f(\zeta)= e^{ih(\alpha)},$
where
$\zeta=e^{i\alpha},\, 0\leq\alpha\leq 2\pi,\, h(0)=h(2\pi),\,
0\leq h(0)<2\pi.$
The real-valued function
$\alpha\rightarrow h(\alpha)$
is smooth. The subgroup
${\cal V}_{0}$ 
of the general procedure is then defined by the equation
\[
\tilde{\cal V}_{0}=T1_{{\cal H}_{0}}\cdot {\cal V}_{0}\cong
T1_{{\cal H}_{0}}\times {\cal V}_{0}
\]
so that for the functions
$U_{f}\in{\cal V}_{0}$
the zero-Fourier coefficient of $h$ vanishes, i.e.
$h_{0}=0$
or
$\int_{0}^{2\pi}h(\alpha)d\alpha =0$,
where
$h(\alpha)=\sum_{n\in Z}h_{n}e^{in\alpha}.\,h(\cdot)$
acts as a multiplication operator $A$ on
$L^{2}(S^{1}).$
Then
$\gamma A\gamma=A$
and
${\cal L}$
is defined as the set of all
$A$
belonging to $f$ with
$U_{f}\in{\cal V}_{0}.$

The first requirement of the last subsection 
is
$PAP^{\bot},P^{\bot}AP\in{\cal L}_{2}({\cal H}_{0}).$
This is true because with
$A=\sum_{n}A_{n}\zeta^{n}$
one obtains
\[
\sum_{n\in Z}\Vert P^{\bot}APe_{n}\Vert^{2}=
\sum_{m=1}^{\infty}m\vert A_{m}\vert^{2}<\infty.
\]
The second requirement of the last subsection,
that ``${\cal V}_{0}$ is abelian" is obvious. 
 The third
requirement says:
$\langle\cdot,\cdot\rangle$
is a scalar product. This is true because of
\[
\mbox{tr}\,(PAP^{\bot}AP)=\sum_{m=1}^{\infty}m\vert A_{m}\vert^{2},
\]
so that
$A_{m}=0$
for
$m\geq 1$
implies
$A_{-m}=\overline{A_{m}}=0,\,m\geq 1.$
Moreover
$A_{0}=0.$
The fourth requirement can be fulfilled choosing
$V_{1}:=U_{f_{1}}$
where, for example,
$f_{1}(\zeta):=\zeta.$
Then
$V_{1}{\cal L}V_{1}^{-1}={\cal L}$
is obvious. Therefore the general approach works and we arrive at the following 

\vspace{0.3cm}
\noindent PROPOSITION 5. {\em Let}
${\cal F}:=C^{\ast}(\Phi(U_{L(T)}))$
{\em be the}
$C^{\ast}$-{\em algebra 
generated by the implementers of the loop group and put}
$\alpha_{\lambda}:=\mbox{Ad}\,\Phi(\lambda)|{\cal F},\,\lambda\in T.$
{\em Then}
$\{{\cal F},\alpha_{T}\}$
{\em is Hilbert,}
${\cal A}\cong C^{\ast}(\Phi(T))\otimes\rho(CCR({\cal L},s)),\,{\cal A}'
\cap {\cal F}={\cal Z},\, {\cal Z}=C^{\ast}(\Phi(T)),\,$
{\em and}
$C(\mbox{spec}\,{\cal Z}\rightarrow T)\cong \mbox{stab}\,{\cal A}.$

\subsection{The case 
$\mbox{spec}\,{\cal Z}=S^{1}$}

We refer to the Weyl-algebra
${\cal B}:=CCR({\cal L},s)$
of the foregoing section, where
$\nu$
is the special automorphism, given by
$\mbox{Ad}\,\Phi(V_{1})$
in the representation
$\rho$.
Recall that
$(\nu^{n},\iota)=\{0\}$
for all
$n\in Z$, $ {\cal B}$
is equipped with a natural net structure w.r.t.
$S^{1}:$
if
$\Delta\subset S^{1}$
is an open interval with
$\mbox{clo}\,\subset S^{1}$
then the assigned $C^{\ast}$-algebra
\[
{\cal B}(\Delta):=C^{\ast}(W(A),\mbox{supp}\,A\subset\Delta)
\]
has the properties

\noindent (I)$\,\Delta_{1}\subset\Delta_{2}\,\mbox{implies}\,
{\cal B}(\Delta_{1})\subset{\cal B}(\Delta_{2}),$

\noindent (II) if $\mbox{clo}\,\Delta_{1}\cap\mbox{clo}\,\Delta_{2}=\emptyset$
\,then ${\cal B}(\Delta_{1}$ and
${\cal B}(\Delta_{2}$
commute elementwise.

\noindent (III) ${\cal B}=\mbox{clo}\{\cup_{\Delta\subset S^{1}}{\cal B}
(\Delta)\}.$

Concerning property (II) recall that the symplectic form $s$ can be rewritten as
\[
s(A,B)=\frac{1}{2\pi}\int_{0}^{2\pi}A(\alpha)B'(\alpha)d\alpha.
\]
Further let
$\delta$
be an automorphism of
$C(S^{1}).$
By $\sigma$ we denote the corresponding homeomorphism of
$S^{1}: (\delta f)(\mu)=f(\sigma(\mu)),\,\mu\in S^{1},\,f\in C(S^{1}).$
Putting
\[
{\cal A}:=C(S^{1})\otimes {\cal B}
\]
and
\[
\kappa :=\delta\otimes \nu
\]
then
$\kappa\in\mbox{aut}\,{\cal A},\,(\kappa^{n},\iota)=\{0\}$
for all
$n\in Z$
follows. According to general extension results (see [BL2] for example) there is
a Hilbert system
$\{{\cal F},\alpha_{T}\}$
w.r.t. the action group $T$ where
${\cal A}\in {\cal F},\, {\cal A}$
the fixed point algebra of
$\alpha_{T}.$
This means there is a unitary
$V\in {\cal F}$
with
$\alpha_{\lambda}V=\lambda V$
for all
$\lambda\in T.$
Obviously,
${\cal Z}:={\cal Z}({\cal A})\cong C(S^{1}).$
Recall that

\begin{equation}
{\cal F}=\mbox{clo}_{\Vert\cdot\Vert}\{\sum_{n\in Z}A_{n}V^{n},\,
A_{n}\in {\cal A},\,\mbox{finite sum}\}.
\end{equation}

\noindent The net structure of
${\cal B}$
can be extended to
${\cal A}$
by
\[
{\cal A}(\Delta):={\cal Z}\otimes {\cal B}(\Delta),\quad \Delta\subset S^{1}.
\]
This means that the elements of
${\cal A}(\Delta)$
are continuous functions
\[
S^{1}\ni\mu\rightarrow B(\mu),\,B(\mu)\in {\cal B}(\Delta)\,\mbox{for all}\,
\mu\in S^{1}.
\]
Furthermore, using (11) the net structure can be extended to
${\cal F}.$

The property
$\mbox{stab}\,{\cal A}\cong C(S^{1}\rightarrow T)$
means that the action of an element $f$ of the group
$C(S^{1}\rightarrow T)$
on the algebra
${\cal F}$
is well-defined if one adopts the point of view to consider $f$ as an element of
$\mbox{stab}\,{\cal A}$
which is justified by the mentioned isomorphism. Recall that the action of
$\beta\in \mbox{stab A}$
is given by
\[\beta (V)=VZ
\]
where
$Z\in{\cal U}({\cal Z})$
is the central unitary, corresponding to the function
$f\in C(S^{1}).$
Then one obtains
\[
\beta (V^{n})=(VZ)^{n}=
\kappa(Z)\kappa^{2}(Z)...\kappa^{n}(Z)V,
\]
or, written as a function on $S^{1}$,
\[
S^{1}\ni\mu\rightarrow
f(\sigma(\mu))f(\sigma^{2}(\mu)...f(\sigma^{n}(\mu))V^{n}.
\]
This leads to the following formula for
$\beta(\sum_{n}A_{n}V^{n}):$
\[
S^{1}\ni\mu\rightarrow
\sum_{n}f(\sigma(\mu))f(\sigma^{2}(\mu))...f(\sigma^{n}(\mu))B_{n}(\mu)V^{n}.
\]
This formula shows that the action of $f$ is compatible with the net structure
of the model.

\vspace{0.3cm}

\noindent REMARK. The example in this subsection
 arises in the study of two dimensional
electrodynamics (the Schwinger model) although in a slightly more complicated
form. In [CW] 
the field algebra of the Schwinger model is constructed.
In this model there is group of so-called `local gauge transformations'
which contains $T$, the global gauge group.
The subalgebra of the field algebra, invariant under local gauge
transformations, is of the form ${\cal A}_0 = {\cal Z}_0\otimes{\cal B}_0$
where ${\cal Z}_0$ is the centre of ${\cal A}_0$.
In [CW] it is further observed that one may pass to a quotient
algebra of ${\cal A}_0$ to obtain the algebra of observables. This
latter algebra
is of the form $C(S^1)\otimes {\cal C}$ where
${\cal C}$ is the algebra of the CCR for a free massive
boson field and hence is a simple algebra. The discussion above
requires no more than simplicity of $\cal B$ and hence applies to
 $C(S^1)\otimes {\cal C}$.
 The presence of the algebra $C(S^1)$ in this tensor product
is a remnant of the local gauge symmetry in the model
in the sense that the generator of this $C(S^1)$ acts in the full
Hilbert space of the model as a gauge transformation. The
interpretation 
in this case is to the existence of a vacuum degeneracy in the
Schwinger model (the so-called $theta$-vacua). Specifically,
the representation of  $C(S^1)\otimes {\cal C}$ which arises
in the Schwinger model from the construction in 
[CW] can be decomposed as a direct integral over the spectrum
of the centre, $S^1$,  into
irreducibles and the GNS cyclic vector in each of these irreducibles has the
interpretation of a vacuum state.
This interpretation is also available in the abstract setting
of Hilbert systems above where the representation of the
algebra $\cal A$ on the Hilbert space for the field algebra
is also decomposable (due to the non-trivial centre).

\section{Acknowledgements}

One of us (H.B.) is grateful for the hospitality at  
the University of Adelaide,
for support by the Australian Research Council,  for the
hospitality at the
Erwin Schroedinger International Institute for Mathematical Physics and the
support by this institute where the paper was prepared finally.
It is a pleasure to thank Professor J. Yngvason for a
discussion on the subject.

\end{document}